\pdfoutput=1 
\documentclass[a4paper,openany, 12pt]{book}

\usepackage[text={15.5cm,23cm},centering]{geometry}

\usepackage{mathptmx}
\usepackage{graphicx}
\usepackage{subfigure}
\usepackage{svg}
\usepackage{color}
\usepackage[colorlinks=true, linkcolor=blue]{hyperref}
\title{European Physical Society Grand Challenges\\
Physics for Society in the Horizon 2050}
\author{EPS Forum Physics and Society}

\begin{document}
\frontmatter
\maketitle
\tableofcontents

\chapter*{Contributors}
 
{\parskip=12pt
\noindent\textbf{Zeki Can Seskir}\\
 Karlsruhe Institute of Technology - ITAS \\
 Karlstraße 11, 76133 Karlsruhe, Germany

\noindent\textbf{Jacob Biamonte}\\
 Skolkovo Institute of Science and Technology \\
 Bolshoy Boulevard 30, bld.~1, Moscow 121205, Russian Federation}

\mainmatter

\chapter{Milestones of research activity in quantum computing}

\subsection*{Chapter Abstract}
We argue that quantum computing underwent an inflection point circa 2017.  Long promised funding materialised which prompted public and private investments around the world.  Techniques from machine learning suddenly influenced central aspects of the field.  On one hand, machine learning was used to emulate quantum systems. On the other hand, quantum algorithms became viewed as a new type of machine learning model (creating the new model of {\it variational} quantum computation).  Here we sketch some milestones which have lead to this inflection point.  We argue that the next inflection point would occur around when practical problems will be first solved by quantum computers.  We anticipate that by 2050 this would have become common place, were the world would still be adjusting to the possibilities brought by quantum computers.\\

{\noindent \textbf{Keywords:} \textit{quantum computing, quantum algorithms, citation analysis, technology forecasting.} }

\subsection*{General Overview}

\subsubsection*{What is quantum computing?}

Today's computers which we know and love---albeit smartphones or the mainframes behind the internet---are all built from billions of transistors.  (A transistor is an electrically controlled switch ultimately the power behind any electronics.) While transistors utilize quantum mechanical effects (such as tunneling: in which an electron can both penetrate and bounce off an energy barrier concurrently), the composite operation of today's computers is purely deterministic or {\it{classical}}.  By classical, we mean {\it{classical mechanics}} which is exactly the physics (a.k.a.~mechanics) we'd anticipate day to day in our lives.  Quantum computers are not intended to always replace classical computers.  Instead, they expected to be a different tool to complement certain types of calculations. We will elaborate shortly.

The term {\it{quantum mechanics}} dates to 1925 in work~\cite{Born1925} by Born and Jordan (in German, {\it{quanten mechanik}}---without the space of course) and comprises the physics governing atomic systems. Quantum mechanics contains principles and rules that appear to contradict the classical mechanics we are so intuitively familiar with. Such counter intuitive phenomena provide new possibilities to store and manipulate (quantum) information. This is exactly what a quantum computer should do. The information must be stored and processed in the matter. You can think of a quantum computer as providing new mechanisms to store, process, and generally manipulate data. Indeed, the ultimate limitation of computational power is given by quantum physics

\subsubsection*{How did quantum computing begin?}

Quantum computing dates back at least to 1979, when the young physicist Paul Benioff~\cite{Benioff1980} at Argonne National Laboratory, USA proposed a quantum mechanical model of computation. Richard Feynman~\cite{Feynman1982} and independently Yuri Manin~\cite{manin} suggested that a quantum computer had the potential to simulate physical processes that a classical computer could not. Such ideas were further formulated and developed in the work of Oxford's David Deutsch~\cite{deutsch1985quantum}---Deutsch formulated a quantum Turing machine and applied a sort of {\it{ anthropic principle}} to the plausible computations allowed by the laws of physics.  Namely, what we now call the Church–Turing–Deutsch principle asserts that a universal (quantum) computing device can simulate any physical process.  (This hypothesis does not give an algorithm but just an assertion that such an algorithm exists).
Yet even the most elementary quantum systems appear impossible to fully emulate using classical computers.  Whereas quantum computers would readily emulate other quantum systems~\cite{Feynman1982,Lloyd1996}.

Early insights into the computational power of quantum computers were based on the assertion that it is impossible to develop an efficient classical algorithm able to accurately emulate quantum systems~\cite{Feynman1982, Lloyd1996}. While not formally proven, ample empirical evidence supports this claim. Yet there are many computational problems where the required computational resources are better understood than simulating physics.   These problems arise in the form of e.g.~the theory of numbers, groups, or properties of graphs.  For decades a small number of researchers worked to understand if one might expect an exponential quantum speedup for problems long studied in computer science.  The early algorithms, such as Deutsch-Jozsa\cite{DJ1992} and Simon's,\cite{Simon1997} solved elegantly contrived problems making prima facie practical merit difficult to envision.

A more practical breakthrough waited until 1994 when Peter Shor proposed an efficient quantum algorithm for factoring integers: this would open the door to decrypt RSA-secured communications~\cite{Shor97}. Shor's algorithm, if executed on a quantum computer, would require exponentially less time than the best known classical factoring algorithms.  Other seminal early findings include Grover's 1996 quantum algorithm~\cite{grover1996fast} which can search through $N$ items in a database in $\sqrt{N}$ steps.  It turns out that Grover's algorithm is provably optimal: it is not possible for a quantum computer to search $N$ unstructured elements any faster than $\sqrt{N}$ steps~\cite{Bennett1997}, which could be significant in practice.  Can these and other quantum algorithms (see Table~\ref{table:quantum}) be realized experimentally~\cite{Aaronson2015}? 

\begin{table}[ht!]
\begin{center}
  \begin{tabular}{|p{6cm}|p{6cm}|}
  \hline
      \textbf{Problem Class} & \textbf{Quantum Complexity} \\
      \hline
      Factorization & Polynomial~\cite{Shor97}\\\hline
      Discrete Log & Polynomial~\cite{Shor97}\\\hline
      Complete Search & $\mathcal{O}(\sqrt N)$~\cite{grover1996fast}\\\hline
      Sparse Linear Systems & Polynomial~\cite{HHL}\\\hline 
      Quantum Simulation & Polynomial~\cite{Lloyd1996} 
 \\ \hline
  \end{tabular} \caption{({\bf Expected Quantum Complexity}) Theory predicts that quantum computers have the potential to rapidly execute several important algorithmic tasks.  Originally developed for the gate model, variational counterparts to Grover's search~\cite{PhysRevA.98.062333}, optimisation (QAOA~\cite{farhi2014quantum}), linear~\cite{bravoprieto2020variational} (and non-linear~\cite{PhysRevA.101.010301}) systems, and quantum simulation (VQE---first proposed in~\cite{Yung2014} and first demonstrated in~\cite{peruzzo2014variational}) have been developed.  Factorization and Discrete Log can readily be mapped to Ising optimisation problems, yet the scaling remains unclear.  Polynomial time variational quantum factoring might instead be accomplished by means of the circuit to variational algorithm mapping~\cite{Bia21}\label{table:quantum}}
  \end{center}
\end{table}

\subsection*{Where is quantum computing heading?}

As of today, billions of dollars of public and private investment are being spent to build quantum computers~\cite{Biamonte2019}.  In October of 2019, Google, in partnership with NASA, performed a quantum sampling task that appears infeasible on any classical computer~\cite{Arute2019}. Late in 2020 Chinese scientists~\cite{Zhong1460} reported results of like-significance.  At the core of these developments is the quantum mechanical bit: the {\it qubit} as first coined by Benjamin Schumacher~\cite{PhysRevA.51.2738} who asserts the term arose out of a 1996 conversation with William Wootters (for qubit, see Table \ref{table:qubit}). Simply, a qubit is a two-state (a.k.a.~binary) quantum system.  

\begin{table}[ht!]
\begin{center}
\begin{tabular}{|l|}
\hline The \textbf{Quantum Bit or Qubit.} \\
\hline A traditional computer operates using bits, 0 and 1.  Each classical \\
register or classical memory must be in a single classical state which is \\
represented by a string of bits (such as 0011 or 0100000111 for example). \\
Quantum computers allow quantum registers and memories to be in multiple \\
states concurrently.  For this reason, quantum computers are often described \\
as enabling parallel processing. \\
\hline
\end{tabular} 
\caption{A brief comparison of traditional bits and quantum bits (qubits).} \label{table:qubit}
\end{center}
\end{table}

There are a variety of physical approaches to creating qubits. For example, a single photon of light can represent two quantum states given by vertical- and horizontal polarization. 
Common qubit realisations include superconducting electronics~\cite{Arute2019}, trapped-ions~\cite{Bruzewicz2019}, and various optical realisations~\cite{RevModPhys.79.135}. Qubits are the building blocks of fully programmable (e.g.~universal) quantum computers: the universal quantum Turing machine's abstraction from physics makes it harder to realize and consequently, this model has fallen from interest~\cite{Benioff1980}. 

Qubits should both be isolated from their surroundings yet also be made to interact. Quantum algorithms are described by {\it quantum circuits}---such circuits depict actions on and interactions between qubits by quantum gates and encompass today's defacto universal model of quantum computation.  In practice, design imperfections and random noise can not be avoided, meaning that the ideal qubit can never exist.  Such noise processes serve to restrict quantum circuit depth. 
Over the last several decades researchers have developed a powerful theory of quantum error correction, proving that qubits need not be perfect to realise high-depth quantum circuits~\cite{Nielsen2009}.  Indeed, quantum error correction proves that a small amount of noise can be tolerated: called the error tolerance threshold~\cite{Shor}.  Hence, according to the laws of physics, nothing fundamentally prevents humans from building a quantum processor capable of executing high-depth quantum circuits.  

A universal quantum computer is assumed be be error tolerant through error correction~\cite{PhysRevA.52.R2493, Georgescu2020}.  Throughout the history of quantum computation, several universal models of quantum computation have been developed and shown to be computationally equivalent to the defacto quantum circuit model~\cite{Nielsen2009}. This includes adiabatic quantum computation~\cite{Farhi472,Aharonov} both discrete and continuous quantum walks~\cite{PhysRevLett.102.180501, Lovett2010}, measurement based quantum computation~\cite{PhysRevLett.97.150504} as well as one of the authors installment proving universality of the variational model~\cite{Bia21}.

Assuming this idealized (universal) setting, several miraculous quantum algorithms have been developed which would offer an advantage over the best-known classical algorithms.  Lower bounding the computational resources required in quantum (and classical) algorithms has however proven extremely difficult: how might we rule out the existence of a better algorithm when the computational power of the class of possible algorithms is not fully understood.

For example, regarding recent quantum adversarial advantage demonstrations:~\cite{Arute2019, Zhong1460}  who is to say that a classical algorithm won't one day be discovered which can replicate the reported sampling task(s)?  This does not imply that such assumptions are not without formal footing.  Elegant methods exist to compare the power of a classical computer to the power of a quantum computer~\cite{Harrow2017}. It does however make the timeline for a practically meaningful quantum computation difficult to predict.  So with all of the dramatic progress, what might we expect from NISQ era quantum processors (NISQ: Noisy Intermediate Scale Quantum Computing~\cite{Preskill2018})? 

\subsection*{Design constraints}

Design constraints limit manufacturing qubits. Working with current design constraints means that we must find ways to utilize imperfect qubits.  The question is if we can still build meaningful systems with imperfect devices.  

Despite outperforming classical computers at an {\it adversarial advantage}~\cite{Arute2019, Zhong1460}, such a demonstration was tailored to favor quantum processors and has unclear practical application.  Even with Moore's law failing~\cite{Waldrop2016}, traditional computing resources are ever-improving and increasingly accessible.  Moreover, the von Neumann architecture has a first-mover advantage:  the entire tool chain, the compilers, algorithms etc.~in use today are tailored towards this architecture.  To instead utilize quantum systems as a computational paradigm represents a dramatic change in how problems must be decomposed, encoded, and compiled.  And in how we think about computing.  Perhaps gleaning lessons from sampling~\cite{Arute2019, Zhong1460}, we must learn to look towards problems that are more amenable to quantum processors, with desirable criteria, such as: 
\begin{enumerate}
    \item[(i)] Problems which bootstrap physical properties of a quantum processor to reduce implementation overhead(s). 
    \item[(ii)] Systems and/or models of computation which offer some inherent tolerance to noise and/or systematic faults.  
    \item[(iii)] Problems which utilize the ability of quantum systems to efficiently represent quantum states of matter---called the {\it memory scaling argument} (see Table \ref{table: memoryscaling}).  
\end{enumerate}

\begin{table}[ht!]
\begin{center}
\begin{tabular}{|l|}
\hline The \textbf{memory scaling argument} asserts that quantum states exist which can \\
not be stored using even the largest classical memories. \\
\hline Early arguments for quantum computing advantage considered an ideal state \\
of interacting qubits, requiring about $2^{{n+1}} \cdot16$ bytes of information to \\
store assuming $32$ bit precision. This reaches $80$ terabytes $(T B)$ at just \\
less than $43$ qubits and $2.2$ petabytes $(P B)$ at just under $47$ e.g - the \\
world's largest memory of the supercomputer Trinity. Hence applications with \\
$\geq 47$ qubits might already outperform classical computers at certain \\
tasks. While this argument didn't account for noise and \\
approximations/compression schemes to reduce required memory, similar \\
arguments are considered valid lines of reasoning today.\\
\hline
\end{tabular} \caption{The memory argument asserts that even the worlds largest computer can not store into its memory all but the simplest quantum states.} \label{table: memoryscaling}
\end{center}
\end{table}

These constraints have led us to what is now called the variational model of quantum computation.  In the absence of error correction, NISQ era quantum computation is focused on quantum circuits that are short enough and with gate fidelity high enough that these short quantum circuits can be executed without quantum error correction, as in the recent quantum supremacy experiments~\cite{Arute2019, Zhong1460}. Herein lies the heart of the variational model: by adjusting parameters in an otherwise fixed quantum circuit, low-depth noisy quantum circuits are pushed to their ultimate use case. 

NISQ circuits typically bootstrap experimentally desirable regularities inline with criteria (i): the gate sequence itself is fixed, while the gate angles can be varied. A classical computer will adjust parameters of a circuit.  Measurements will be used to calculate a cost function and the process will be iterate (see Table \ref{table:zoo} for some examples).

\begin{table}[ht!]
\begin{center}
  \begin{tabular}{|p{5cm}|p{4.5cm}|p{5cm}|}
  \hline
      \textbf{Problem Hamiltonian} & {\bf Finding Ground Energy} (Classical / Quantum) & {\bf Calculating State Energy}  (Classical / Quantum) \\
      \hline
      {\sc 1-Local Hamiltonian} & Polynomial & Polynomial\\\hline
      {\sc 2-Local Ising} & $^\star$Exp & Polynomial \\\hline
      Electronic Structure & $^\star$Exp & $^\star$Exp / Polynomial\\\hline
      ZZXX Model & $^\star$Exp & $^\star$Exp / $^\star$Polynomial\\\hline 
  \end{tabular} \caption{({\bf Hamiltonian Complexity Micro Zoo}) Anticipated computational resources to determine ground state energy and calculate energy relative to a state.  `Restricted Ising' denotes problems known to be in {\sf P}. The upper star ($^\star$) denotes expected and not formally proven conjectures. Electronic Structure problem instances have constant maximum size so are assumed to be in {\sf BQP} whereas the ZZXX model admits a {\sf QMA}-complete ground state energy decision problem}\label{table:zoo}
  \end{center}
\end{table}

The prospects of the variational model are limited by the computational overhead of outer-loop optimisation.  This requires significant classical computing resources. Variational model proposes some alternatives to this (for a further comparison of these models, please see Table \ref{table:comp}

\begin{table}[ht!]
\begin{center}
\begin{tabular}{|p{7.2cm}|p{7.2cm}|} 
\hline
{\bf Variational} & {\bf Traditional}  \\ \hline
\begin{itemize}
    \item  Agnostic to systematic errors
    \item  Tightly connects hardware with software to overcome hardware constraints
    \item  Optimizes short depth circuits for optimal use 
    \item  Emulates Hamiltonians by local measurements
    \item Outer loop optimization can require significant classical computing resources 
    \item Coherence time and error rates limit circuit depth
\end{itemize} & 
\begin{itemize}
    \item Intuitive and familiar, textbook quantum algorithms adhere to the circuit model
    \item Theoretical analysis, including complexity, has largely been proven possible
    \item  Impossible to execute all but the shortest circuits (smallest examples) with current hardware
    \item  Ignores hardware constraints and susceptible to both systematic and random errors
\end{itemize}\\
\hline
\end{tabular} \caption{Comparison of standard gate model quantum computation versus variational quantum computation.  Variational quantum computation trains short quantum circuits to reach their maximum use case yet requires significant classical coprocessing to train these quantum circuits.} \label{table:comp}
\end{center}
\end{table}

\subsection*{How did quantum computing develop prior to 2017?}

A turning point in the development of quantum computation appears around 2017.  At this point, several long-promised large funding programs began such as the European Quantum Flagship and the American National Quantum Initiative Act (this happened around the world and was in the Billions of USD).  Most national investments appear to keep a country competitive in technological development. There are many initiatives around the world adding up to more than 20 billion USD committed public funding. In addition, many private companies also invested dramatically around this time.

Meanwhile, quantum computation was merged with machine learning in two different ways (see the review \cite{BWN+17}).  Firstly, quantum circuits can be trained variationally.  In other words, quantum circuits can be viewed as machine learning models.  Secondly, machine learning can be applied to a host of problems faced in building and emulating quantum systems.  These two facts encouraged the tech industry participation in quantum computing research and development.  (In fact, a new model of comptuation was developed and proven to be universal in \cite{Bia21} by one of the authors.) 

While those working in the field might readily agree that things have rapidly developed since around 2017, putting data behind this claim is the focus of this section.  Hence, we will quantitatively describe the levels of activity before 2017, especially focusing on the field of quantum algorithms. 

We have utilized two data sources for this section, one for academic publications and one for patent publications; Web of Science owned by Clarivate Analytics PLC, and the Cipher platform owned by Aistemos Ltd. However, our readers can use the same queries\footnote{For quantum software and algorithms: (("quantum machine learning") OR ("qml" AND "quantum") OR ("quantum approximate optimization") OR ("vqe" AND "quantum") OR ("variational quantum eigen*") OR ("quantum algorithm*") OR ("quantum software") OR (“quantum Machine Learning”) OR (”Classical-quantum Hybrid Algorithm*”) OR (“quantum PCA”) OR (”quantum SVM”) OR (”variational Circuit*” AND "quantum") OR (”quantum Anneal*” AND "algorithm*") OR (”quantum Enhanced Kernel Method*”) OR (”quantum Deep Learning”) OR (”quantum Matrix Inversion”) OR ("quantum embed*") OR ("quantum neural") OR ("quantum perceptron") OR ("quantum tensor network*")) \newline
For artificial intelligence: (("machine learning") OR ("artificial intelligence") OR ("neural network*")) \newline
Date: June 11th 2021} to reproduce our results. To generate our queries we have used the previous ones created by previous publications in the literature \cite{Pande2020,Seskir2021}, and personal expertise.


\begin{figure}[ht!]
    \centering
    \includegraphics[scale=0.8]{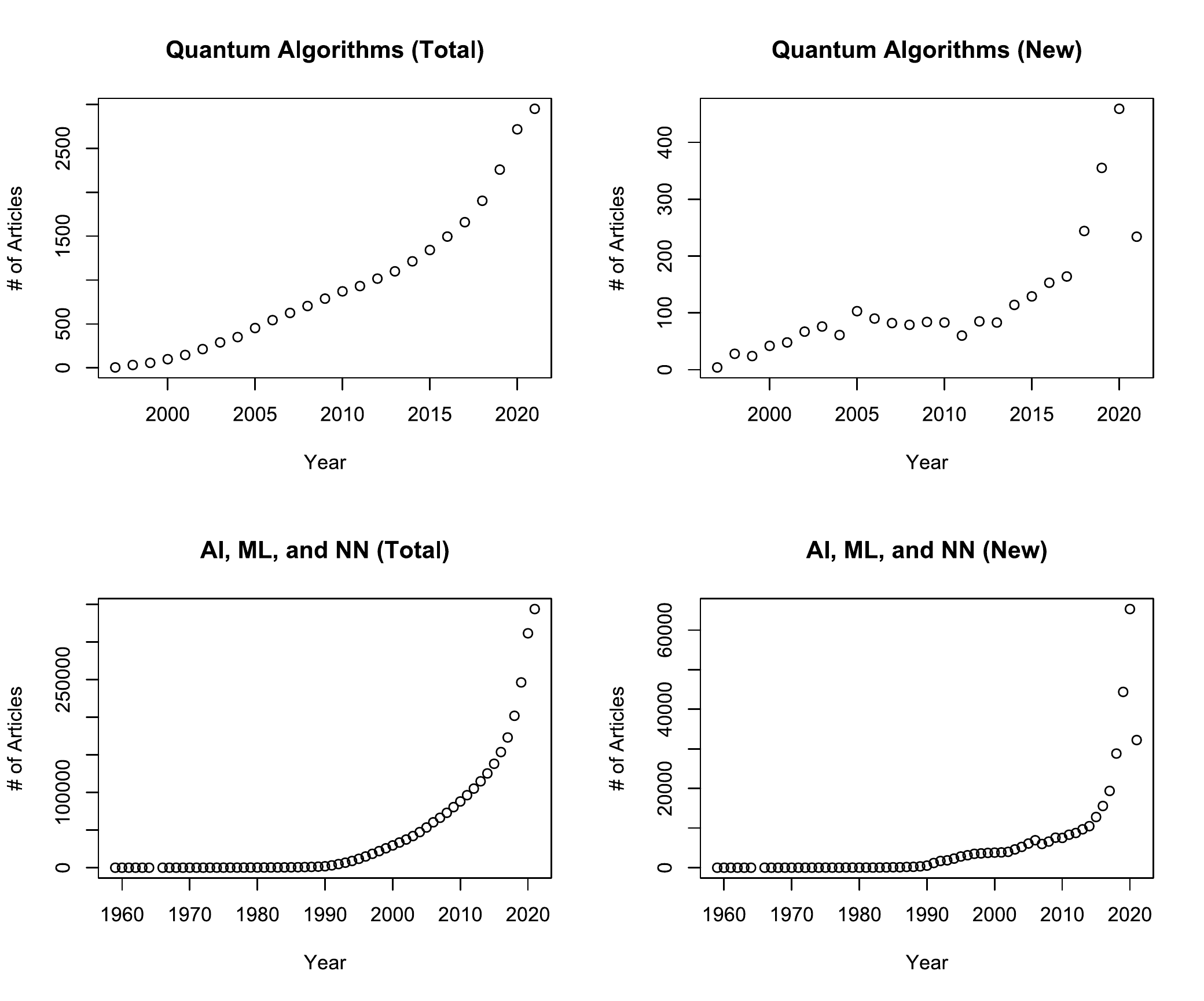}
    \caption{The numbers of new and total published articles in Quantum Algorithms and A.I. fields.}
    \label{fig_publications_qalg}
\end{figure}

To build our study, we first consider Figure \ref{fig_publications_qalg}.  We see in the upper left-hand pane, the growth of quantum algorithms articles.  We do see what appears to be an increase between 2015 and 2020. By total, we mean the entire sum up until that point. By new, we mean the number of articles in a given year.  For the left top page of Figure \ref{fig_publications_qalg}, we again see a jump between 2015 and 2020.

Additionally, we ran the publication data set through a software tool for constructing and visualizing bibliometric networks (VOSViewer \cite{vosviewer}) to run author keyword clustering, which resulted in the following Figure \ref{keywords_map}. In this figure, each node represents an author keyword, size of the node is correlated with how many times the keyword appears in the data set, connections between nodes represent co-occurrences of those keywords, and the color scale represents the average year of the keyword in the literature. 


\begin{figure}[ht!]
    \includegraphics[width=1.15\textwidth,keepaspectratio]{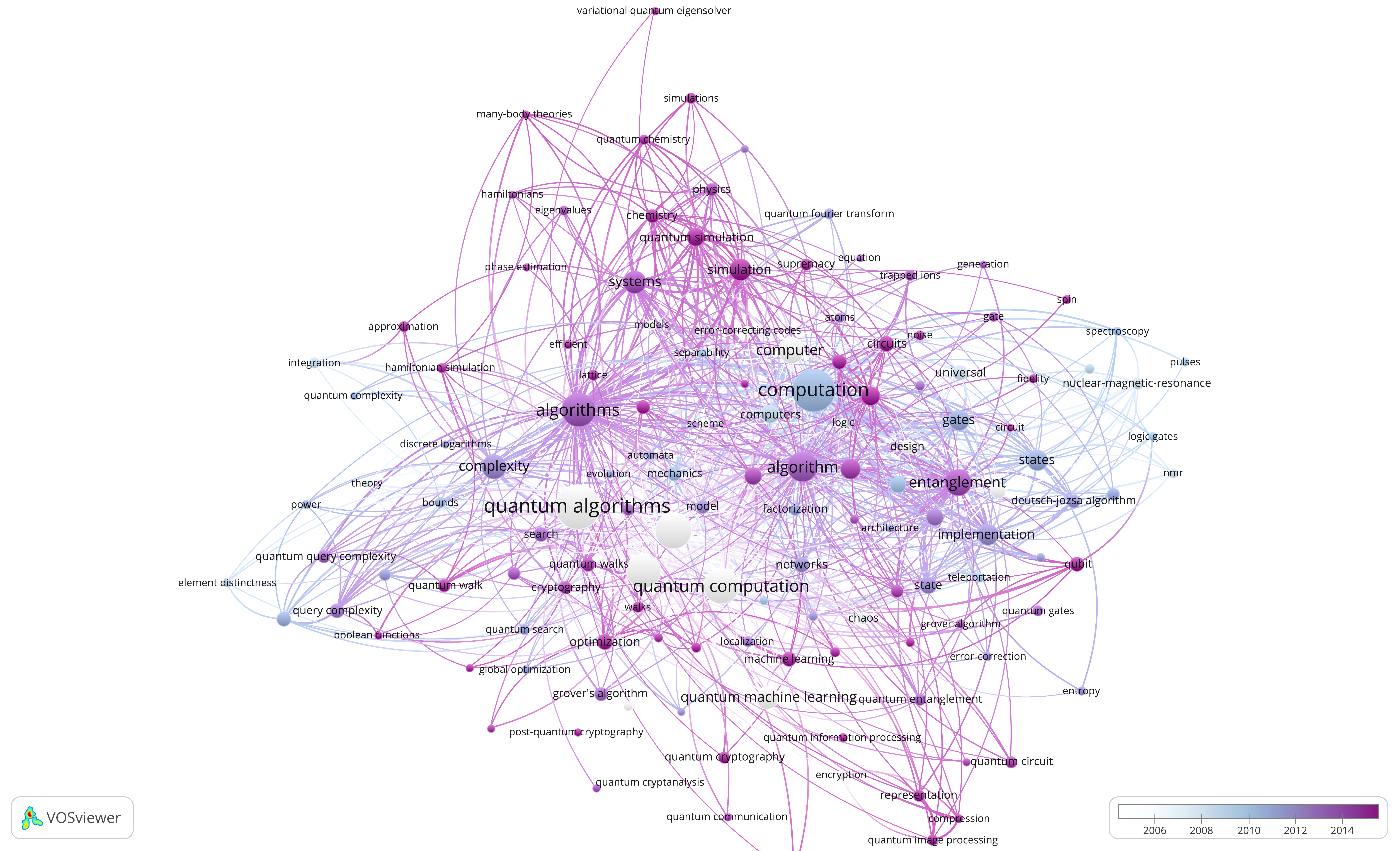}
    \caption{Overlay visualization of author keywords used in the academic articles in time.}
    \label{keywords_map}
\end{figure}

Here, one can notice the following point. Keywords such as \textit{quantum algorithms, quantum computation, quantum computer} are mainly the keywords utilized by older literature, which in recent years are replaced with more field specific terms like \textit{quantum chemistry, variational quantum eigensolver} and \textit{quantum image processing}. This indicates an evolution of the literature into partially distinct lines of research, which are more developed topics in terms of maturity, compared to earlier keywords utilized in the literature.


\begin{figure}[ht!]
    \centering
    \includegraphics[width=1.15\textwidth,keepaspectratio]{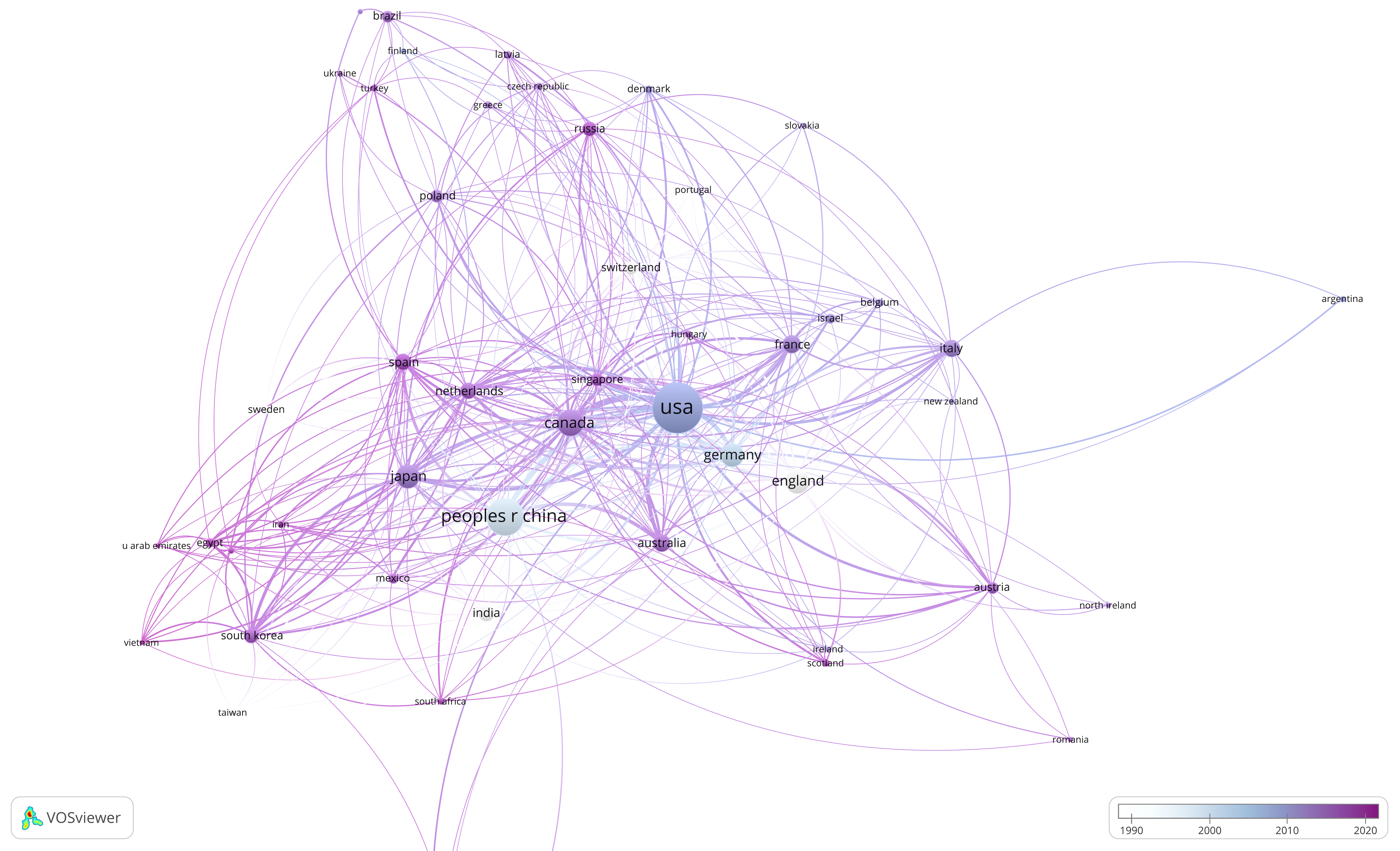}
    \caption{Overlay visualization of countries of origin of academic articles in time.}
    \label{countries_map}
\end{figure}

We also ran a similar analysis for the collaboration between countries (Figure \ref{countries_map}). The country level data is associated with the affiliations of authors, and this map only represents the academic literature created by cross-country collaboration. This visual reveals that, although countries like England, Germany, China, and the US are located in the center of the collaboration network, a considerable number of new countries joined this network in recent years (hence, represented by dark purple in the figure).

\begin{figure}[ht!]
    \centering
    \includegraphics[scale=0.8]{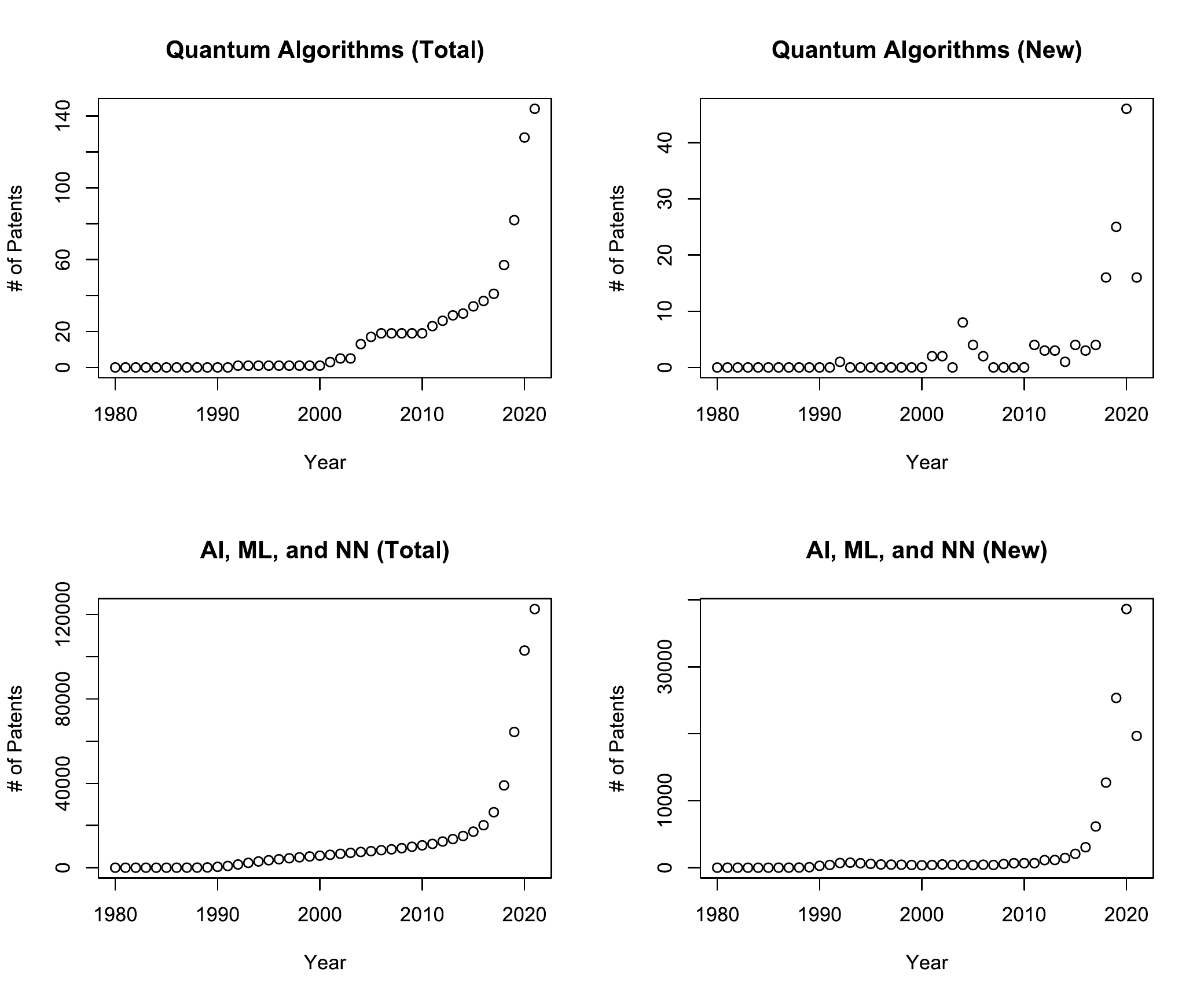}
    \caption{The numbers of new and total patents in Quantum Algorithms and A.I. fields.}
    \label{fig_patents_qalg}
\end{figure}

In terms of numbers, the patent literature reveals similar results to academic publications.  We see a rise in the early 2000s in the top left pane in Figure \ref{fig_patents_qalg}.  We again see a jump before 2020.  That sharp jump is the turning point we are discussing.  The new and total number reflect this.  We see similar jumps in the AI fields at a much larger scale (bottom two figures). Comparing the trends in Figure \ref{fig_patents_qalg} with the Figure \ref{fig_publications_qalg}, it is clear to see that scientific interest in both fields have been more steady compared to commercial interest until around 2016-2017, and then both starts in a strongly upward trend.


\begin{figure}[ht!]
    \centering
    \includegraphics[scale=0.8]{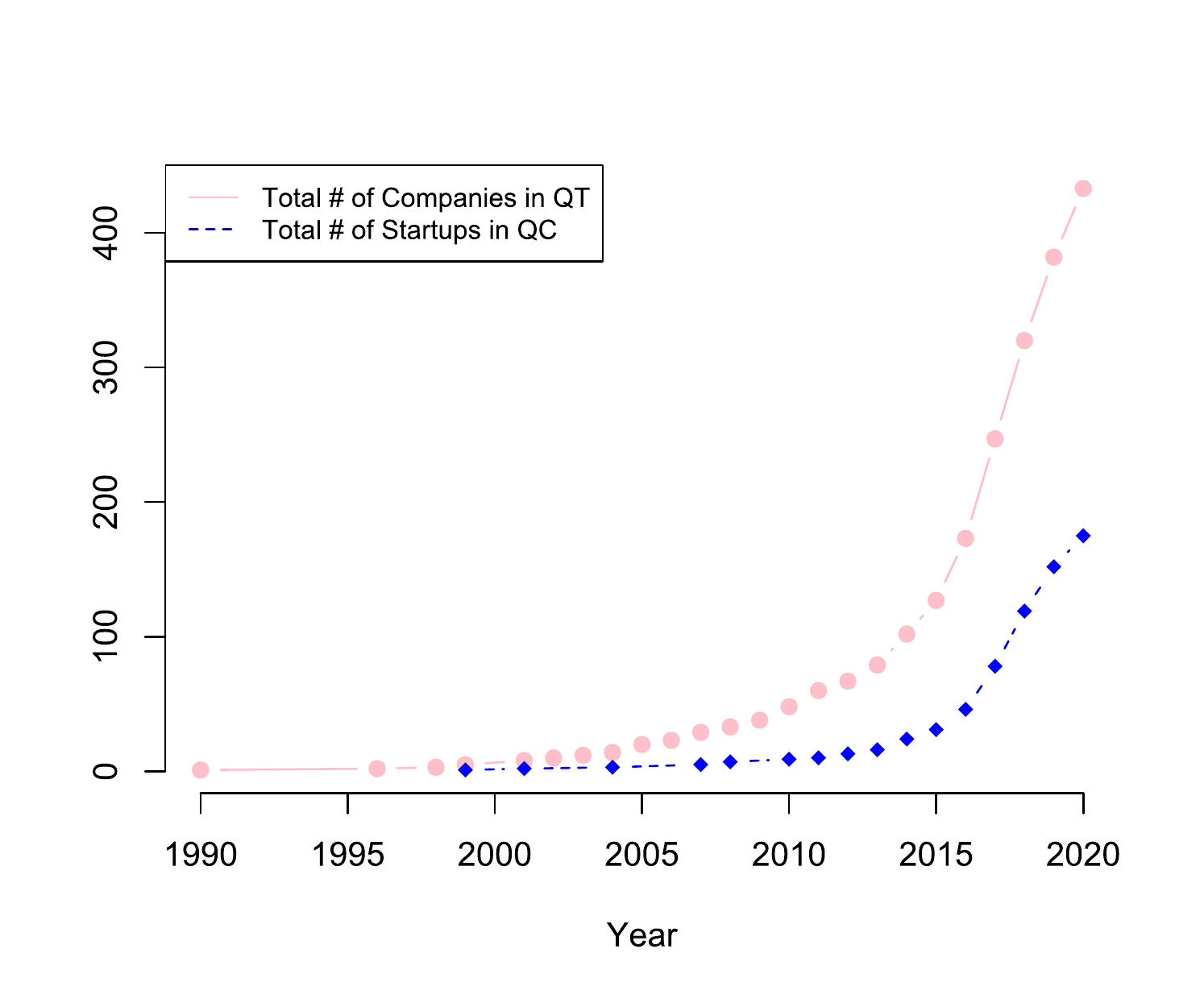}
    \caption{Total number of companies operating in quantum technologies and startups in quantum computing in years.}
    \label{number_of_companies}
\end{figure}

The above exploration of the academic and patent landscapes revealed several important insights into the current state of the field. First of all, comparison between Artificial Intelligence and Quantum Computing in terms of commercialization is clearly an overstatement in terms of patents. Using the queries given, we were able to find 122,609 patent families in the A.I.~field versus 144 in quantum algorithms. Similarly, there were 343,808 articles in A.I. and 2,951 in quantum algorithms. These represent two-to-three orders of magnitude differences in publications and patents between these fields. In this sense, quantum algorithms can be related to the early 1990s when compared to A.I., which might provide some insight into how the algorithm and quantum computing might evolve in the coming decades.  

Finally, to see how this activity has been translating in the entrepreneurial realm, we gathered a list of companies \footnote{This list was collected by us manually and contains 439 companies as of June 2021. We used open access resources like Crunchbase, LinkedIn, The Quantum Insider, and Quantum Computing Report.} in quantum technologies (QT), and identified the startups related to quantum computing (QC) in them, to compare them in Figure \ref{number_of_companies}. It should be noted that other fields under QT (such as quantum sensing and quantum communication/cryptography) have also been gaining popularity in the recent years. From the figure, we can see that startups in QC is a relatively new phenomenon compared to companies in QT, but both have been rapidly increasing in number during the last five years.

In summary, some historical differences and similarities between these fields can be seen in Figures \ref{fig_patents_qalg} and \ref{fig_publications_qalg}. One clear difference is of the scales, as there are orders of magnitude between the fields, and the second difference is that from early 1990 there has been a steady increase in the total number of patents obtained in the field of A.I. compared to the almost zero activity in the field of quantum algorithms except a brief and short lived interest in mid-2000s. One clear similarity is the sudden spike in late 2010s, especially after 2016-2017, which is also evident in Figure \ref{number_of_companies}. The reason for this can be attributed to long promised funding materialising which in return prompted public and private investments around the world. Origins of some of the public funding schemes can be attributed to fear of missing out for countries with existing scientific investments in the field (Figure \ref{countries_map}). Some can be explained by public demonstrations of IBM's and Google's superconducting quantum processors, signaling to the public (and investors) that quantum computing is becoming into the realm of calculable risk from Knightian uncertainty \cite{watkinsknight1922}. Regardless of the specifics, it is clear that both the number of new academic articles and new patents per year, and the number of startups in QC have increased significantly starting from 2016-2017.

This inflection point was not gone unnoticed in the ELSA (Ethical, Legal and Social Aspects) field as well, and one of the first special issues on QT titled \textit{The Societal Impact of the Emerging Quantum Technologies} was published in the Ethics and Information Technology journal \footnote{ISSN	1388-1957 (print) 1572-8439 (web)}. Topics like access to quantum technologies \cite{DiVincenzo2017}, the impact of quantum computing on the future of scientific computing \cite{Moller2017}, responsible research and innovation in quantum technologies \cite{Coenen2017}, and the potential impact of quantum computers on society \cite{Wolf2017} were discussed in the special issue. Following 2017, there has been a growing interest in the societal impact of quantum technologies (and quantum computing in particular). As of 2021, many researchers and commercial actors in the field have been calling the community to action regarding quantum ethics, which is a clear indication that there is a strong belief in the community that this technology will play a huge role in our future, and should be developed ethically to avoid any undesired consequences while taking this quantum leap.

\section*{What is the next `turning point' in the development of quantum computing?}

For us at least it’s hard to predict technology. We can assume everything works and imagine best-case scenarios. We can assume nothing works and imagine a sort of quantum winter. In reality, it’s likely the case that the changes ahead are impossible to envision today. 

Quantum computing violates no known laws of physics. It, therefore, is thought of as perhaps the world's most challenging engineering problem. So when will we engineer such devices? Currently, we have seen dramatic progress. Progress so dramatic that it would be impossible for those working in the field to have imagined even 5 years ago. Perhaps this means that even five years from today will be difficult to predict. We still believe in five years from now that quantum computers will be in the early stages of development and still lack error correction and other essential features to realize all their potentials. 

Saying that we do think that a quantum future is inevitable. This is the natural progression of technology. This is probably the ultimate limit of computers, perhaps until we can harness new powers in the cosmos. Humankind’s trajectory is set on a quantum course. Companies such as Google, IBM, and the like are in this for research and long-term prospects. 

When will we see another inflection point? It’s hard to tell. The saying goes that knowledge begets knowledge. And so development always seems to go increasingly faster. But the next jump might have to wait until practical problems of commercial value are regularly solved. This should take place perhaps even around 2050. Yet by then, we do imagine that this technology would have already changed the world in ways we can’t hope but predict now

\noindent {\bf Acknowledgements.} JB acknowledges support from the project, Leading Research Center on Quantum Computing (Agreement No.~014/20). 


{
\bibliographystyle{unsrt}
\bibliography{iopebook}
}
\end{document}